\documentclass[conference]{IEEEtran}
\IEEEoverridecommandlockouts
\usepackage{subcaption}
\usepackage{cite}
\usepackage{amsmath,amssymb,amsfonts}
\usepackage{algorithmic}
\usepackage{graphicx}
\usepackage{textcomp}
\usepackage{xcolor}
\def\BibTeX{{\rm B\kern-.05em{\sc i\kern-.025em b}\kern-.08em
    T\kern-.1667em\lower.7ex\hbox{E}\kern-.125emX}}
\begin{document}

\title{UAV Swarming for Air-Ground ISAC via Cross-Region Cooperation 
}

\author{\IEEEauthorblockA{Linghui Miao, Shijian Gao\\
The Hong Kong University of Science and Technology (Guangzhou), China \\
}

}

\maketitle

\begin{abstract}
To serve the volumetric air-ground space, uncrewed aerial vehicles (UAVs) are urgently needed. Yet, relying on them for integrated sensing and communication (ISAC) introduces two key challenges: 1) dynamic and imbalanced ground communication demand, and 2) limited observation diversity for sensing. To address these issues, a cross-region cooperative framework is designed to coordinate UAV swarms. Specifically, a service-driven regional partitioning scheme is proposed to support traffic-aware UAV communication, and an adaptive handshaking mechanism is introduced to improve cooperative sensing accuracy by mitigating residual inter-region phase errors with controlled synchronization overhead. Based on these designs, a region-level multi-agent proximal policy optimization (MAPPO) framework with centralized training and decentralized execution (CTDE) is developed for cross-region cooperative decision-making. Simulation results demonstrate that the proposed method achieves a communication quality-of-service (QoS) of approximately 90\% and reduces the Cramér-Rao bound (CRB) by about 45\% compared to conventional baselines.

\end{abstract}


\section{Introduction}

To support communication and sensing services in the volumetric air-ground
space, uncrewed aerial vehicles (UAVs) are emerging as key aerial nodes owing
to their flexible deployment, controllable three-dimensional mobility, and
favorable line-of-sight air-ground links \cite{gao2026iscc}. These features are well aligned with integrated sensing and communication (ISAC), supporting data transmission and target sensing over shared wireless resources \cite{liu2022isac}. Therefore, UAV-enabled ISAC has become a promising paradigm for low-altitude and mobile networks
\cite{tang2025lae,yang2025som}.

Building on the mobility of UAV platforms, existing UAV-enabled ISAC studies
have evolved from single-UAV designs to multi-UAV cooperation. Single-UAV-assisted ISAC systems have been studied through joint trajectory, reconfigurable intelligent surface (RIS) phase-shift, and resource-allocation optimization \cite{amhaz2025uav,huroon2025aavris}. However, a single UAV is inherently limited by its coverage range, onboard resources, and single-view sensing capability, which makes it difficult to scale to large-area or multi-user ISAC scenarios. To overcome these limitations, recent research has shifted toward multi-UAV cooperation, where properly exploiting collective spatial degrees of freedom can enhance both communication coverage and sensing accuracy in large-area and complex ISAC scenarios. Existing works have considered joint trajectory and resource allocation \cite{qin2023deep,pan2024cooperative}, user association and beam management \cite{zhang2024joint}, and learning-based cooperative control for scalable multi-UAV decision-making \cite{gao2024marl}.

While existing multi-UAV ISAC studies have advanced post-deployment UAV control, practical air-ground scenarios are also affected by spatially imbalanced and time-varying ground traffic demand. Related cluster-level coordination studies on vehicular task offloading \cite{hou2024hierarchical}, space-air-ground task scheduling \cite{wang2026cluster}, and UAV mobility management \cite{meer2026dynamic} have shown that regional load heterogeneity is an important factor in network-level resource coordination. In UAV-enabled ISAC, such regional demand heterogeneity becomes coupled with sensing, since UAVs serving ground vehicles also act as sensing nodes. Meanwhile, relying only on UAVs near the target may restrict observation diversity, especially when these UAVs also need to maintain local communication services. Cooperative perception \cite{wang2024cooperative}, distributed sensing \cite{meng2025cooperative}, and network-level ISAC coordination \cite{meng2024networklevel} have demonstrated the benefits of multi-node cooperation for sensing enhancement. However, coherent sensing among distributed UAVs is sensitive to residual phase errors, and synchronization is recognized as a key factor in cooperative ISAC systems \cite{liu2025cooperative}.

Motivated by these limitations, this paper proposes a cross-region cooperative
UAV swarming framework for air-ground ISAC. In practical air-ground scenarios,
ground vehicles are dynamically and unevenly distributed, leading to
time-varying and imbalanced communication service demands across different
regions. At the same time, accurate aerial-target sensing generally requires sufficient observation diversity among distributed UAVs to establish favorable observation geometries. These two tasks are tightly coupled and often conflicting, since UAV positions and transmit powers affect both the quality of ground-user service and the effectiveness of sensing information acquisition. To address these issues, a service-driven regional partitioning scheme is first designed to  divide the coverage area according to traffic-demand hotspots, enabling UAVs to better adapt their communication services to localized user demands. For sensing, cross-region UAV coordination is established for target localization, where an adaptive handshaking mechanism is introduced to mitigate residual inter-region phase errors with controlled synchronization overhead and energy cost. Based on these designs, a region-level multi-agent proximal policy optimization (MAPPO) framework with centralized training and decentralized execution (CTDE) is developed for cross-region cooperative decision-making, aiming to balance communication quality-of-service (QoS), sensing accuracy, and UAV energy consumption.

\section{System Model}

\begin{figure}[!t]
\centering
\includegraphics[width=0.95\linewidth]{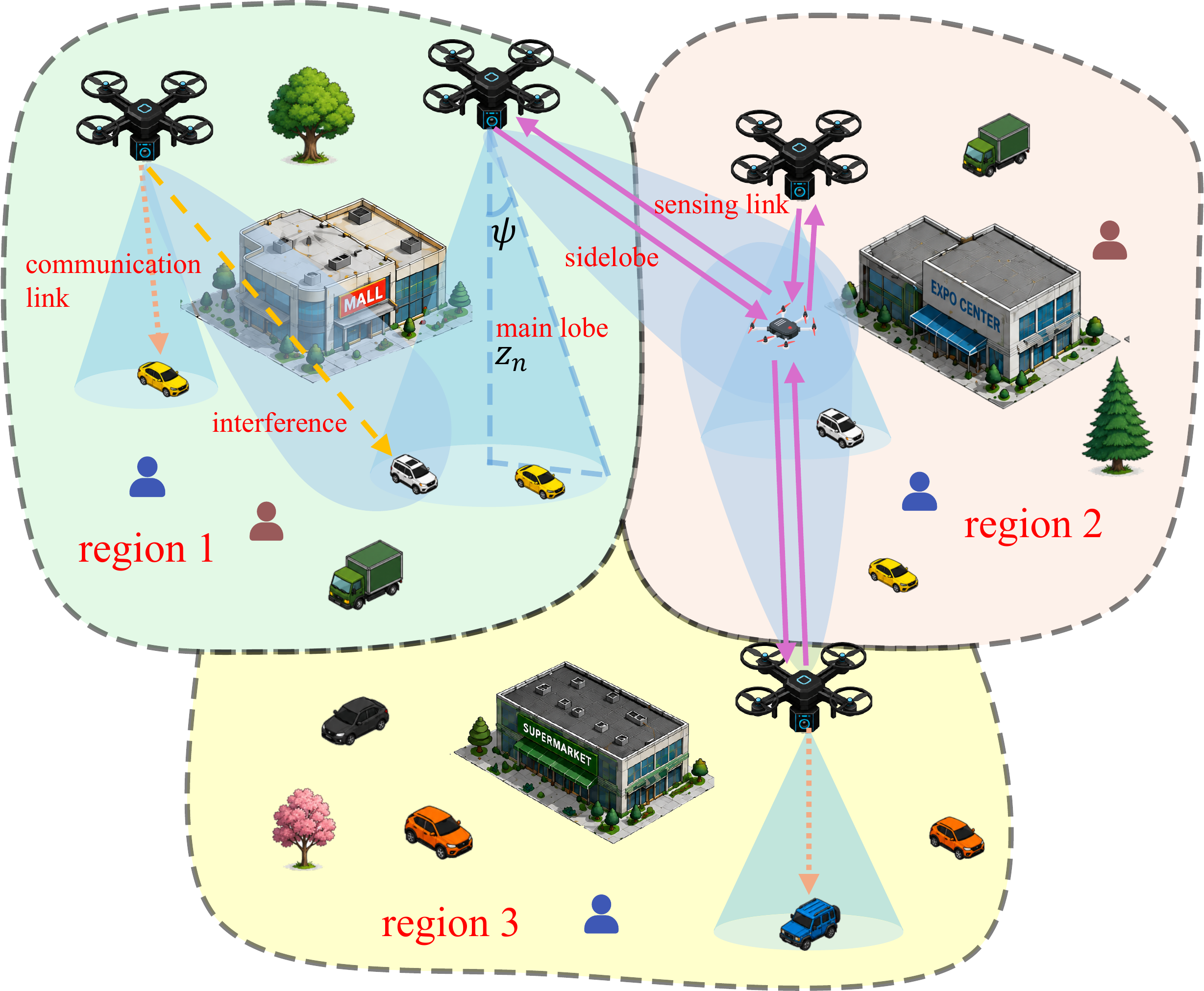}
\caption{UAV-enabled ISAC system model with regional partitioning.}
\label{fig:system_model}
\end{figure}

As illustrated in Fig.~\ref{fig:system_model}, we consider an air-ground ISAC
system supported by a UAV swarm, where UAVs serve ground vehicles and
cooperatively sense an aerial target. Each UAV is equipped with a vertically
downward directional antenna, and both main-lobe and side-lobe links are
considered. Side-lobe links may contribute to cooperative sensing or cause
inter-region communication interference. For tractability, each UAV beam is
assumed to affect only its own service region and neighboring regions. The service area is partitioned into $M$ service-driven regions by Voronoi
partitioning, where $\mathcal M=\{1,\ldots,M\}$ denotes the region set. The
region centers correspond to representative service hotspots rather than
instantaneous vehicle positions. Each region is associated with a subset of UAVs for local service, while the regional adjacency structure enables neighboring UAVs to assist aerial-target sensing. Inter-region handshaking is further used to reduce synchronization errors for coherent sensing.

The system operates over discrete time slots
$t\in\mathcal T=\{1,\ldots,T\}$ with slot duration $\Delta t$. Under the adopted
slotted-time approximation, the positions of UAVs, vehicles, and the aerial
target are treated as quasi-static within each slot and evolve across slots.
Let $\mathcal N=\{1,\ldots,N\}$ and $\mathcal U=\{1,\ldots,U\}$ denote the
overall sets of UAVs and ground vehicles, respectively. For region $m\in\mathcal M$, let $\mathcal N_m\subseteq\mathcal N$ denote the
set of UAVs assigned to region $m$, and let
$\mathcal U_m(t)\subseteq\mathcal U$ denote the set of ground vehicles
associated with region $m$ at slot $t$. The positions of UAV
$n$, vehicle $u$, and the aerial target are denoted by
$\mathbf l_n(t)=[x_n(t),y_n(t),z_n(t)]^T$,
$\mathbf q_u(t)=[x_u(t),y_u(t),0]^T$, and
$\mathbf p(t)=[x_p(t),y_p(t),z_p(t)]^T$, respectively. To characterize cross-region interaction, let $\widetilde{\mathcal M}_m$
denote the set of neighboring regions of region $m$. The neighboring-region UAV set associated with region $m$ is defined as
$\widetilde{\mathcal N}_m=\bigcup_{\ell\in\widetilde{\mathcal M}_m}
\mathcal N_\ell$. When the aerial target is located in region $m$, the
effective sensing UAV set is given by
\begin{equation}
\mathcal C_m
=
\mathcal N_m\cup \widetilde{\mathcal N}_m .
\end{equation}

Based on the directional antenna model introduced above, $\Theta$ denotes the
half-power beamwidth of each UAV antenna. For a vehicle $u\in\mathcal U_m(t)$,
the off-boresight angle between the downward beam axis of UAV $n$ and the link
toward vehicle $u$ is denoted by $\psi_{n,u}(t)$, where
\begin{equation}
\cos\psi_{n,u}(t)
=
\frac{
\mathbf e_z^T\left(\mathbf l_n(t)-\mathbf q_u(t)\right)
}{
\|\mathbf l_n(t)-\mathbf q_u(t)\|
},
\label{eq:beam_angle_comm}
\end{equation}
and $\mathbf e_z=[0,0,1]^T$. The antenna gain between UAV $n$ and vehicle $u$
is modeled as
\begin{equation}
G_{n,u}(t)=
\begin{cases}
G_{\mathrm{main}}, 
& n\in\mathcal C_m,\ \psi_{n,u}(t)\le \Theta/2,\\
G_{\mathrm{side}}, 
& n\in\mathcal C_m,\ \psi_{n,u}(t)> \Theta/2,\\
0,
& n\notin\mathcal C_m.
\end{cases}
\label{eq:antenna_gain_comm}
\end{equation}
where $G_{\mathrm{main}}$ and $G_{\mathrm{side}}$ denote the main-lobe and
side-lobe gains, respectively.

UAVs move in response to ground-vehicle dynamics, while transmit power and handshaking frequency further affect signal strength, energy consumption, and synchronization overhead. Therefore, UAV mobility, transmit power, and inter-region handshaking need to be jointly controlled to balance regional communication service, sensing accuracy, and energy consumption.

\section{Utility Functions for UAV-Swarm ISAC}
Based on the system model described above, this section characterizes the
utility functions for UAV-swarm ISAC, including communication service utility,
cross-region cooperative sensing utility, and energy-aware utility, and then
formulates the joint control problem.

\subsection{Communication Service Utility}

Let $P_n(t)$ and $R_{n,u}(t)$ denote the transmit power of UAV $n$ and its
distance to vehicle $u$, respectively. Under the line-of-sight (LoS) propagation assumption, the communication link follows the free-space path-loss model with carrier wavelength $\lambda\approx c/f_c$, where $c$ is the speed of light and $f_c$ is the carrier frequency. If UAV $n$ serves $k_n(t)>0$ vehicles, its bandwidth and transmit power are equally allocated, i.e., $B/k_n(t)$ and $P_n(t)/k_n(t)$ for each served vehicle. Vehicles served by the same UAV use orthogonal frequency resources, while UAVs in neighboring regions may reuse the same frequency resources and cause interference. Therefore, for UAV $n\in\mathcal N_m$ and vehicle $u\in\mathcal U_m(t)$, the communication signal-to-interference-plus-noise ratio (SINR) is given by
\begin{equation}
\resizebox{0.91\linewidth}{!}{$
\displaystyle
\gamma_{n,u}^{\mathrm c}(t)
=
\frac{
\frac{P_n(t)}{k_n(t)}
G_{n,u}(t)
\left(
\frac{\lambda}{4\pi R_{n,u}(t)}
\right)^2
}{
N_0\frac{B}{k_n(t)}
+
\sum\limits_{j\in
\mathcal C_m\setminus\{n\}}
\frac{P_j(t)}{k_j(t)}
G_{j,u}(t)
\left(
\frac{\lambda}{4\pi R_{j,u}(t)}
\right)^2
},
$}
\label{eq:sinr_comm}
\end{equation}
where $N_0$ denotes the additive white Gaussian noise (AWGN) power spectral
density, $B$ denotes the system bandwidth, and $k_j(t)$ is the number of
vehicles served by interfering UAV $j$.

Accordingly, the achievable rate of vehicle $u$ served by UAV $n$ is given by
\begin{equation}
R_{n,u}^{\mathrm c}(t)
=
\frac{B}{k_n(t)}
\log_2\!\left(
1+\gamma_{n,u}^{\mathrm c}(t)
\right).
\label{eq:rate_comm}
\end{equation}
Each vehicle $u\in\mathcal U_m(t)$ is associated with the UAV
$n\in\mathcal N_m$ that provides the largest SINR. For each UAV, the associated vehicles are then sorted in descending order
of SINR. Due to equal bandwidth sharing, the maximum number of vehicles that UAV $n$ can serve while satisfying $R_{\mathrm{th}}$ is denoted by $K_n(t)$. Let $\chi_u(t)\in\{0,1\}$ denote the service satisfaction indicator of vehicle $u$, where $\chi_u(t)=1$ if vehicle $u$ is admitted by its associated UAV and satisfies the minimum rate requirement, and $\chi_u(t)=0$ otherwise. The communication QoS of region $m$ is defined as the satisfied service ratio:
\begin{equation}
Q_m(t)
=
\frac{1}{|\mathcal U_m(t)|}
\sum_{u\in\mathcal U_m(t)}
\chi_u(t).
\label{eq:communication_qos}
\end{equation}

\subsection{Cross-Region Cooperative Sensing Utility}

We next formulate the cooperative sensing utility for the aerial target $p$. Assume
that the target is located in region $m$ at slot $t$. The UAVs in region $m$
and its neighboring regions participate in cooperative sensing. For UAV $n\in\mathcal C_m$, the antenna gain from UAV $n$ to the aerial target is denoted by $G_{n,p}(t)$, which follows the same directional antenna gain model as \eqref{eq:antenna_gain_comm}. The sensing signal-to-noise ratio (SNR) of UAV $n$ is modeled according to the monostatic radar link budget as
$\gamma_n^{\mathrm s}(t)=P_n(t)\big(G_{n,p}(t)\big)^2\lambda^2\sigma_{\mathrm{RCS}}\big/\big[(4\pi)^3\big(R_{n,p}^{\mathrm s}(t)\big)^4N_0B\big]$, where $\sigma_{\mathrm{RCS}}$ denotes the radar cross section (RCS) of the aerial target, and $R_{n,p}^{\mathrm s}(t)$ denotes the distance between UAV $n$ and
the aerial target at slot $t$.

Cross-region coherent sensing is affected by residual phase errors among
different regions. Local phase synchronization within each region is assumed to
be maintained, and the target-located region $m$ is taken as the phase-reference
region. For each neighboring region $\ell\in\widetilde{\mathcal M}_m$, the
inter-region phase-error variance with respect to region $m$ is expressed as
\begin{equation}
\resizebox{0.91\linewidth}{!}{$
\displaystyle
\sigma_{\epsilon,\ell}^{2}(t)
=
(2\pi f_c)^2\sigma_{\tau,\ell}^{2}(t)
+
(2\pi T_{\mathrm c})^2\sigma_{f,\ell}^{2}(t)
+
\sigma_{\mathrm{pn},\ell}^{2}(t)
+
\sigma_{\mathrm{cal},\ell}^{2}(t),
$}
\label{eq:expanded_phase_variance}
\end{equation}
where $\sigma_{\tau,\ell}^{2}(t)$, $\sigma_{f,\ell}^{2}(t)$,
$\sigma_{\mathrm{pn},\ell}^{2}(t)$, and $\sigma_{\mathrm{cal},\ell}^{2}(t)$
denote the variances of residual timing offset, carrier-frequency offset,
oscillator phase noise, and calibration error, respectively.

Inter-region handshaking periodically updates timing, frequency, and
phase-reference information among neighboring regions. The residual phase error
contains a non-trackable component caused by hardware impairments, estimation
noise, quantization error, and oscillator phase noise, as well as a trackable
component caused by timing offset, carrier-frequency offset, phase drift, and
calibration latency. The former forms a residual error floor, whereas the latter
decreases as the handshaking frequency increases. Thus, the residual
inter-region phase-error variance is modeled as
\begin{equation}
\sigma_{\epsilon,\ell}^{2}(t)
=
\sigma_{\epsilon,\ell,0}^{2}
+
\frac{\kappa_{\ell}}{f_{\mathrm H}(t)},
\quad
\ell\in\widetilde{\mathcal M}_m,
\label{eq:residual_phase_variance}
\end{equation}
where $\sigma_{\epsilon,\ell,0}^{2}$ denotes the residual phase-error floor that
cannot be further reduced by increasing the handshaking frequency, and
$\kappa_{\ell}$ summarizes the handshaking-sensitive synchronization uncertainty
caused by residual timing offset, carrier-frequency offset, slowly varying phase
drift, and calibration latency.

The residual inter-region phase errors affect the coherent sensing gain. For the
target-located region $m$, the coherence factor is set to one, i.e.,
$\eta_m(t)=1$. For each neighboring region
$\ell\in\widetilde{\mathcal M}_m$, the coherence factor is modeled as
$\eta_\ell(t)=\exp[-\sigma_{\epsilon,\ell}^{2}(t)/2]$, which decreases with the
residual phase-error variance. By accumulating the information contributions from UAVs in $\mathcal C_m$, the Fisher information matrix (FIM) for target localization is given by
\begin{equation}
\mathbf J^{\mathrm s}(t)
=
\alpha_{\mathrm{coh}}(t)
\sum_{n\in\mathcal C_m}
\iota_n^{\mathrm s}(t)
\mathbf a_n(t)\mathbf a_n^T(t),
\label{eq:coherent_sensing_fim}
\end{equation}
where $\alpha_{\mathrm{coh}}(t)$ is the coherent combining gain determined by
the inter-region coherence factors, $\iota_n^{\mathrm s}(t)$ is the sensing
information weight, and $\mathbf a_n(t)$ is the unit direction vector from UAV
$n$ to the target. The target localization accuracy is evaluated by the Cramér--Rao bound (CRB):
\begin{equation}
\Phi_{\mathrm{CRB}}(t)
=
\operatorname{Tr}
\left[
\left(
\mathbf J^{\mathrm s}(t)
\right)^{-1}
\right].
\label{eq:sensing_crb}
\end{equation}

\subsection{UAV-Swarm Energy Utility}

Since UAVs need to adjust their positions according to the time-varying vehicle distribution and sensing geometry, propulsion energy should be considered in
addition to transmission energy \cite{wang2025novel}. Moreover, inter-region handshaking among UAVs is required to reduce synchronization errors for cross-region coherent sensing, which introduces additional handshaking energy consumption.

For UAV $n$, let $\mathbf v_n(t)=[v_{n,x}(t),v_{n,y}(t),v_{n,z}(t)]^T$ denote its
velocity at slot $t$. $P_n^{\mathrm{fly}}(t)$ is the propulsion power, so the propulsion energy of UAV $n$ in slot $t$ can be expressed as $E_n^{\mathrm{fly}}(t)
=P_n^{\mathrm{fly}}(t)\Delta t$. Since each UAV transmits a unified ISAC signal, the transmission energy of UAV $n$ is modeled as $E_n^{\mathrm{tx}}(t)=P_n(t)\Delta t$.
In addition, cross-region coherent sensing requires inter-region handshaking to
update timing, frequency, and phase-reference information among participating regions. The handshake frequency $f_{\mathrm H}(t)$ reduces the residual inter-region phase uncertainty, but it also consumes additional energy for signaling exchange. The handshaking energy is modeled as
\begin{equation}
E^{\mathrm H}(t)
=
\zeta_{\mathrm H} f_{\mathrm H}(t)\Delta t,
\label{eq:handshake_energy}
\end{equation}
where $\zeta_{\mathrm H}$ is the energy coefficient that converts the
handshaking frequency into synchronization-related power. 

\subsection{Problem Formulation}

For region $m$, the UAV velocity, transmit power, and handshaking frequency
are jointly optimized to improve regional communication QoS, reduce the
sensing CRB, and limit energy consumption. Let $\omega_{\mathrm c}$,
$\omega_{\mathrm s}$, and $\omega_{\mathrm e}$ denote the weights of
communication QoS, sensing accuracy, and energy consumption, respectively.
The binary variable $\alpha_m(t)\in\{0,1\}$ indicates whether region $m$
participates in the cross-region cooperative sensing task at slot $t$, where
$\alpha_m(t)=1$ if region $m$ participates in sensing and $\alpha_m(t)=0$
otherwise. In addition, $\eta_{\mathrm{CRB}}$ and $E_{\max}$ are normalization
constants for the sensing CRB and energy consumption, respectively. Then, the optimization problem is formulated as
\begin{subequations}
\label{problem_control}
\begin{align}
\textbf{P1:}\quad
&\scalebox{0.88}{$\displaystyle
\begin{aligned}
&\underset{\substack{\mathbf v_n(t),\,P_n(t),\\ f_{\mathrm H}(t)}}{\max}
\quad
\sum_{t\in\mathcal T}
\Bigg[
\omega_{\mathrm c}Q_m(t)
-
\omega_{\mathrm s}\alpha_m(t)
\frac{\log\!\left(1+\Phi_{\mathrm{CRB}}(t)\right)}{\eta_{\mathrm{CRB}}}
\\
&\hspace{1.0cm}
-
\omega_{\mathrm e}
\frac{\sum_{n\in\mathcal N_m}
\left(
E_n^{\mathrm{fly}}(t)
+
E_n^{\mathrm{tx}}(t)
\right)+\alpha_m(t) E^{\mathrm H}(t)}{E_{\max}}
\Bigg]
\end{aligned}
$}
\notag
\\
\mathrm{s.t.}\quad
&
\left\|\mathbf v_n(t)\right\|
\le
V_{\max},
\quad
\forall n\in\mathcal N_m,
\label{eq:control_speed}
\\
&
0
\le
P_n(t)
\le
P_{\max},
\quad
\forall n\in\mathcal N_m,
\label{eq:control_power}
\\
&
f_{\mathrm H}^{\min}
\le
f_{\mathrm H}(t)
\le
f_{\mathrm H}^{\max},
\label{eq:control_handshake}
\\
&
z_{\min}
\le
z_n(t)
\le
z_{\max},
\quad
\forall n\in\mathcal N_m,
\label{eq:control_altitude}
\\
&
\mathbf l_n^{xy}(t)
\in
\Omega_m,
\quad
\forall n\in\mathcal N_m,
\label{eq:control_region_boundary}
\\
&
\scalebox{0.95}{$\displaystyle
\left\|
\mathbf l_n(t)-\mathbf l_j(t)
\right\|
\ge
d_{\min},
\quad
\forall n,j\in\mathcal N_m,\ n\neq j,
$}
\label{eq:control_safety_distance}
\end{align}
\end{subequations}
where the constraint \eqref{eq:control_speed} restricts the UAV velocity, while \eqref{eq:control_power} and \eqref{eq:control_handshake} impose the feasible ranges of the transmit power and the handshaking frequency, respectively. Constraint \eqref{eq:control_altitude} keeps UAVs within the allowable altitude range. Constraint
\eqref{eq:control_region_boundary} ensures that the horizontal projection of each UAV remains inside its assigned service region, and \eqref{eq:control_safety_distance} guarantees the minimum safety distance between UAVs.

Problem \textbf{P1} is difficult to solve with conventional optimization methods due to its long-term objective, coupled decision variables, and non-convex constraints. The dynamic vehicle distribution further makes repeated real-time optimization impractical. Therefore, reinforcement learning is adopted to learn long-term control policies from environment interactions. Since the joint action space grows rapidly with the number of UAVs and each UAV makes decisions based on local observations, a multi-agent reinforcement learning (MARL) framework is employed for scalable cooperative decision-making.

\section{Cross-Region Cooperation via MARL}

This section presents the learning-based design for UAV-swarm cross-region cooperation. The Markov decision process is first formulated, followed by the cross-region cooperation mechanism and the MAPPO-based policy optimization procedure under the CTDE paradigm.

\subsection{Region-Level Markov Decision Process}

At each time slot $t$, each service region is treated as a local
decision-making unit. The UAVs in $\mathcal N_m$ act as cooperative agents
to serve local vehicles and, when required, contribute to cross-region
cooperative sensing. The decision process of region $m$ is modeled as a
partially observable Markov decision process, characterized by $\left(
\mathcal S_m,\mathcal O_m,\mathcal A_m,r_m,\mathcal P,\gamma
\right)$, where $\mathcal S_m$ is the centralized state space, $\mathcal O_m$ is the local
observation space, $\mathcal A_m$ is the continuous action space, $r_m$ is the
regional reward, $\mathcal P$ denotes the state transition probability, and
$\gamma\in(0,1)$ is the discount factor.

For region $m$, the centralized state used by the critic is denoted by
$\mathbf s_m(t)\in\mathcal S_m$ and is constructed as
\begin{equation}
\mathbf s_m(t)
=
\left\{
\mathbf s_m^{\mathrm{uav}}(t),
\mathbf s_m^{\mathrm{nbr}}(t),
\mathbf s_m^{\mathrm{veh}}(t),
\mathbf s_m^{\mathrm{tar}}(t),
\mathbf s_m^{\mathrm{sync}}(t)
\right\},
\label{eq:centralized_state}
\end{equation}
where the state includes intra-region UAV states, neighboring-region UAV states,
regional vehicle-distribution information, target-related information, and
synchronization/sensing feedback such as $f_{\mathrm H}(t)$, $\alpha_m(t)$,
$\Phi_{\mathrm{CRB}}(t)$, and the marginal CRB utility. For UAV $n\in\mathcal N_m$, the local observation used by the actor is denoted by $\mathbf o_n(t)\in\mathcal O_m$ and is given by
\begin{equation}
\mathbf o_n(t)
=
\left\{
\mathbf o_n^{\mathrm{self}}(t),
\mathbf o_n^{\mathrm{veh}}(t),
\mathbf o_n^{\mathrm{nbr}}(t),
\mathbf o_n^{\mathrm{tar}}(t),
\mathbf o_n^{\mathrm{fb}}(t)
\right\},
\label{eq:local_observation}
\end{equation}
which contains the UAV's self state, local vehicle information, neighboring-UAV
information, target-related observation, and communication/sensing feedback. Each UAV outputs a continuous action vector:
\begin{equation}
\mathbf a_n(t)
=
\left\{
a_{n,x}(t),
a_{n,y}(t),
a_{n,z}(t),
a_{n,\mathrm H}(t),
a_{n,P}(t)
\right\},
\label{eq:action_vector}
\end{equation}
where the first three dimensions control UAV mobility, $a_{n,P}(t)$ controls
the transmit power, and $a_{n,\mathrm H}(t)$ represents the local handshaking
preference of UAV $n$. The regional reward follows the objective in \textbf{P1}. For region $m$, the reward at slot $t$ is given by
\begin{equation}
\scalebox{0.90}{$\displaystyle
\begin{aligned}
r_m(t)
=&\ 
-\Pi_m(t)
+
\omega_{\mathrm c}Q_m(t)
-
\omega_{\mathrm s}\alpha_m(t)
\frac{
\log\!\left(1+\Phi_{\mathrm{CRB}}(t)\right)
}{
\eta_{\mathrm{CRB}}
}
\\
&
-
\omega_{\mathrm e}
\frac{\sum_{n\in\mathcal N_m}
\left(
E_n^{\mathrm{fly}}(t)
+
E_n^{\mathrm{tx}}(t)
\right)+\alpha_m(t) E^{\mathrm H}(t)}{
E_{\max}
}
,
\end{aligned}
$}
\label{eq:regional_reward}
\end{equation}
where $\Pi_m(t)$ denotes the penalty for violating constraints, including
boundary violations and unsafe UAV distances.

\subsection{Cross-Region Cooperation Mechanism}

Based on the region-level actions defined above, we describe how cross-region cooperation is implemented in the proposed UAV-swarm cooperation framework. At each time slot, each region first performs local communication control for its own vehicles. When the aerial target is located in region $m$, the UAVs in the effective
sensing set $\mathcal C_m$ participate in cooperative sensing. Thus, cross-region cooperation is implemented by allowing UAVs in neighboring regions to jointly adjust their mobility and transmit power for the sensing task, while still maintaining their local communication service. Specifically, UAVs in $\mathcal N_m$ provide the basic sensing observations
from the target-located region, while UAVs in $\widetilde{\mathcal N}_m$  provide additional sensing observations from neighboring regions. These UAVs act in a decentralized manner based on their local observations, but their actions jointly determine the sensing geometry, the inter-region interference, and the regional communication QoS. The sensing feedback, including the global CRB and the marginal CRB contribution of each participating region, is then used to guide subsequent cooperative decisions.

To enable coherent sensing across regions, the participating UAVs further perform inter-region handshaking to reduce residual synchronization errors. Since handshaking is shared by the cooperative sensing process, UAV-specific handshaking frequencies are not used. Instead, each UAV outputs a local
handshaking preference $a_{n,\mathrm H}(t)$, and these preferences are aggregated into a common handshaking action:
\begin{equation}
a_{\mathrm H}^{\mathrm g}(t)
=
\sum_{i\in\{m\}\cup\widetilde{\mathcal M}_m}
\beta_i^{\mathrm H}(t)
\frac{1}{|\mathcal N_i|}
\sum_{n\in\mathcal N_i}
a_{n,\mathrm H}(t),
\label{eq:global_handshake_action}
\end{equation}
where $\beta_i^{\mathrm H}(t)$ is the normalized aggregation weight of region $i$, and a larger weight is assigned to a region with a larger marginal CRB contribution. The aggregated action $a_{\mathrm H}^{\mathrm g}(t)$ is linearly
mapped from $[-1,1]$ to $[f_{\mathrm H}^{\min},f_{\mathrm H}^{\max}]$ to obtain the shared handshaking frequency $f_{\mathrm H}(t)$.

\subsection{MAPPO-based Policy Optimization}

Based on the regional reward in \eqref{eq:regional_reward}, MAPPO learns
long-term control policies for UAV mobility, transmit power, and handshaking
preference under the CTDE paradigm. The policy of each UAV is represented by an
actor network $\pi_\theta(\mathbf a_n(t)|\mathbf o_n(t))$, which maps the local
observation to a continuous action distribution. Meanwhile, the centralized
critic $V_\psi(\mathbf s_m(t))$ estimates the value function using the regional
centralized state. The temporal-difference error is defined as
\begin{equation}
\delta_t
=
r_m(t)
+
\gamma V_\psi(\mathbf s_m(t+1))
-
V_\psi(\mathbf s_m(t)),
\label{eq:td_error}
\end{equation}
where $\gamma\in(0,1)$ is the discount factor. The advantage function is
estimated using generalized advantage estimation (GAE):
\begin{equation}
\hat A_t
=
\sum_{l=0}^{T-t-1}
(\gamma\lambda_{\mathrm{GAE}})^l
\delta_{t+l},
\label{eq:gae}
\end{equation}
where $\lambda_{\mathrm{GAE}}$ is the GAE parameter. The actor is updated using the clipped surrogate objective:
\begin{equation}
\scalebox{0.83}{$
\displaystyle
\mathcal L_{\pi}(\theta)
=
\mathbb E_t
\left[
\min\left(
\rho_t(\theta)\hat A_t,
\mathrm{clip}\left(
\rho_t(\theta),
1-\epsilon_{\mathrm{ppo}},
1+\epsilon_{\mathrm{ppo}}
\right)\hat A_t
\right)
\right],
$}
\label{eq:ppo_actor_loss}
\end{equation}
where $\rho_t(\theta)=
\pi_\theta(\mathbf a_n(t)|\mathbf o_n(t))/
\pi_{\theta_{\mathrm{old}}}(\mathbf a_n(t)|\mathbf o_n(t))$
is the probability ratio between the current and old policies, and
$\epsilon_{\mathrm{ppo}}$ is the clipping parameter. The critic is trained by minimizing the value loss:
\begin{equation}
\mathcal L_V(\psi)
=
\mathbb E_t
\left[
\left(
V_\psi(\mathbf s_m(t))
-
\hat R_t
\right)^2
\right],
\label{eq:critic_loss}
\end{equation}
where $\hat R_t$ denotes the discounted return.

\section{Simulations}

\begin{figure}[!t]
\centering
\includegraphics[width=0.95\linewidth]{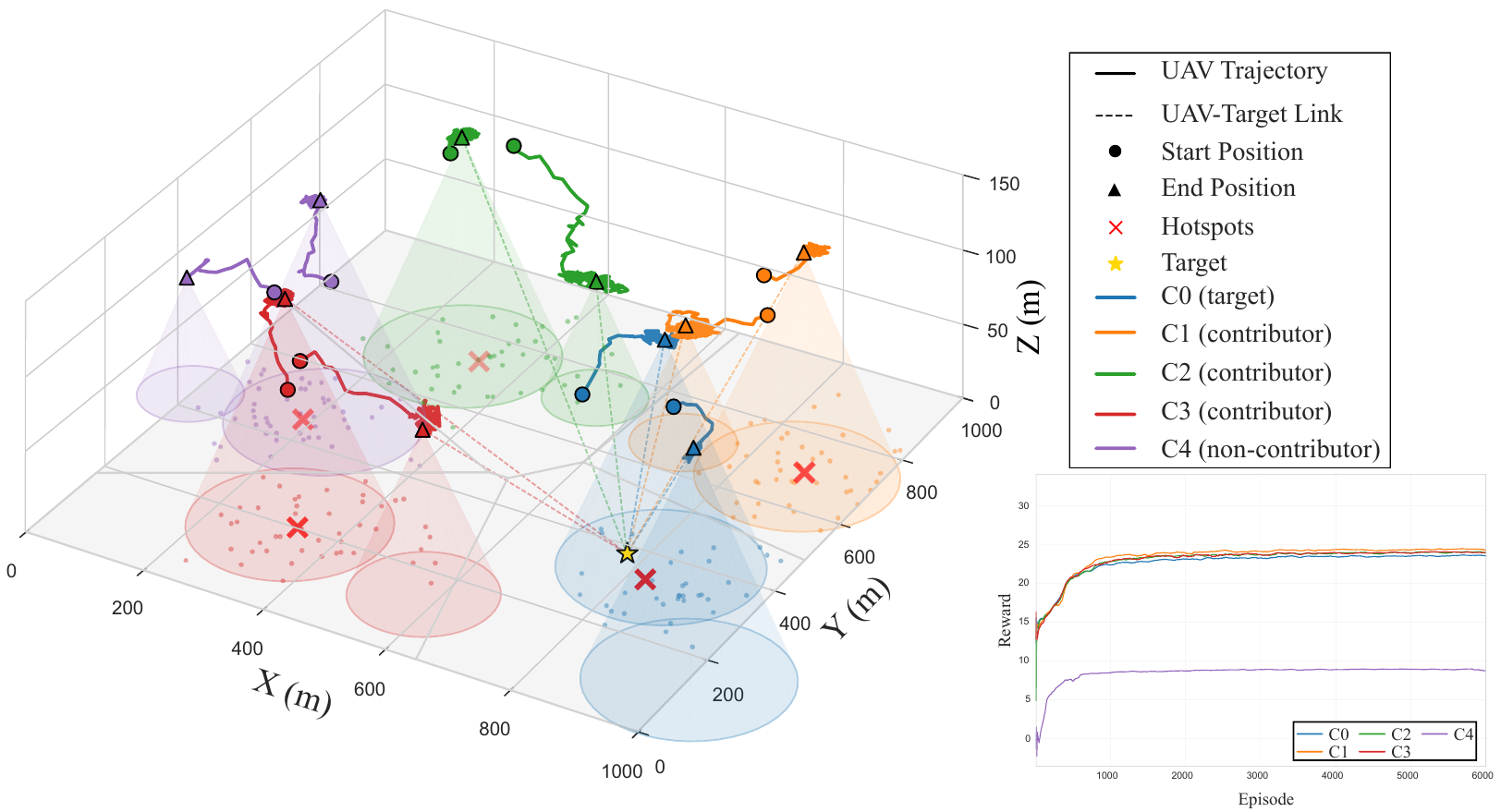}
\caption{3-D UAV trajectories under cross-region cooperation.}
\label{fig:trajectory_3d}
\end{figure}

Simulations are conducted to evaluate the proposed region-level CTDE
MAPPO method. The service area is a $1000\,\mathrm{m}\times1000\,\mathrm{m}$ square region
partitioned into $M=5$ service-driven regions. Each region is assigned two UAVs
to support regional vehicular communication and cross-region cooperative
sensing. The UAV altitude is constrained within $[80,150]\,\mathrm{m}$, and the
maximum speed is $15\,\mathrm{m/s}$. Fig.~\ref{fig:trajectory_3d} illustrates the learned three-dimensional UAV
trajectories. The learned policy does not simply drive
all UAVs toward the target, but coordinates UAV mobility according to both local
vehicle service and cross-region sensing geometry. The graph in the lower right
corner further shows the reward convergence of each region.

\begin{figure}[!t]
\centering
\includegraphics[width=0.85\linewidth]{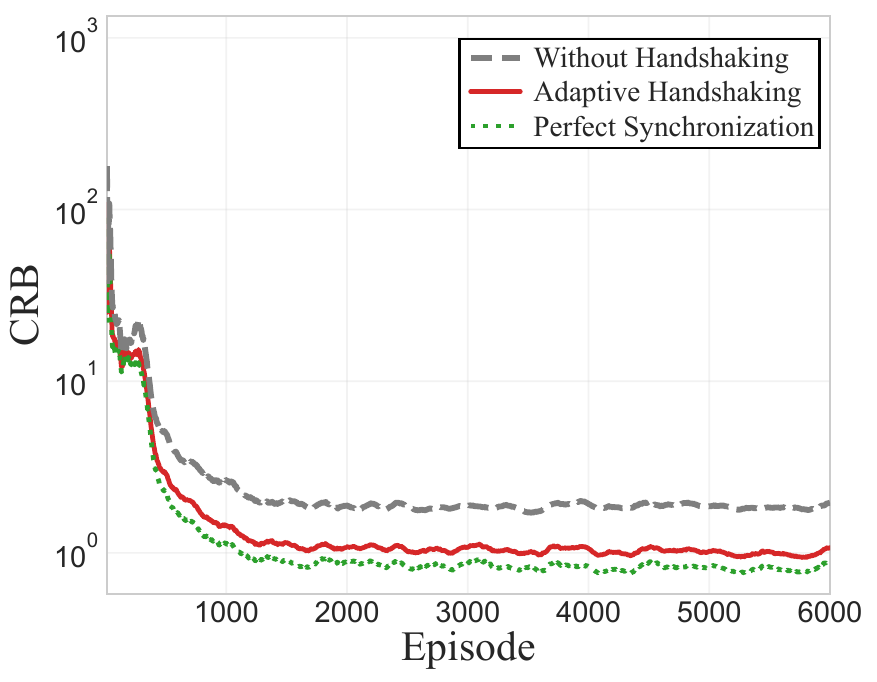}
\caption{CRB comparison under different synchronization settings.}
\label{fig:fh_ablation_crb}
\end{figure}

To evaluate the effectiveness of the handshaking mechanism, an ablation study is
conducted under three synchronization settings: Without Handshaking, Adaptive Handshaking, and Perfect Synchronization. Without Handshaking removes
inter-region synchronization enhancement, Adaptive Handshaking uses
the learned handshaking frequency, and Perfect Synchronization
serves as an ideal benchmark. As shown in Fig.~\ref{fig:fh_ablation_crb}, the Adaptive Handshaking scheme achieves a lower CRB than the Without Handshaking case and approaches the Perfect Synchronization benchmark. In the converged stage, Adaptive Handshaking reduces the CRB by about 45\% compared with the Without Handshaking case. This verifies that the adaptive handshaking mechanism can effectively improve cross-region coherent sensing accuracy.

\begin{figure}[!t]
\centering
\begin{subfigure}{0.85\linewidth}
    \centering
    \includegraphics[width=\linewidth]{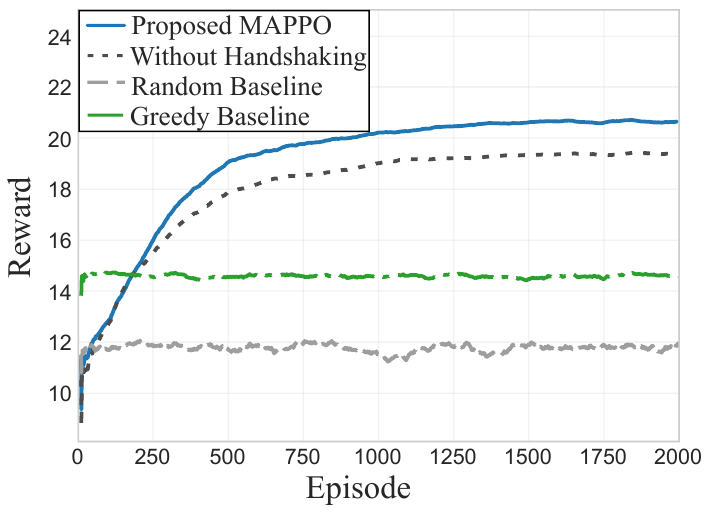}
    \caption{Reward convergence.}
    \label{avg_reward}
\end{subfigure}
\vspace{0.4em}
\begin{subfigure}{0.85\linewidth}
    \centering
    \includegraphics[width=\linewidth]{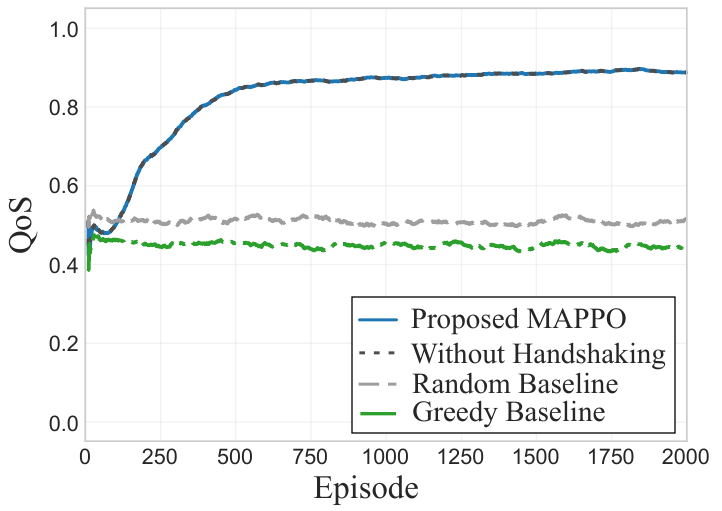}
    \caption{QoS convergence.}
    \label{global_qos}
\end{subfigure}
\caption{Performance comparison under different schemes.}
\label{fig:reward_qos_comparison}
\end{figure}

As shown in Fig.~\ref{fig:reward_qos_comparison}, the proposed MAPPO method is
compared with three benchmarks. Random Baseline selects actions randomly, and Greedy Baseline makes myopic decisions by selecting the action that maximizes the immediate reward at each step. Fig.~\ref{fig:reward_qos_comparison}(a) reports the reward averaged over all service regions, while Fig.~\ref{fig:reward_qos_comparison}(b) reports the global communication QoS over the entire service area. The proposed MAPPO method achieves the highest converged reward and rapidly improves the global QoS to approximately 89\%. The QoS curves of MAPPO and Without Handshaking are the same because the handshaking frequency has no effect on communication QoS. In addition, the Greedy Baseline achieves the lowest observed QoS because its myopic step-wise decisions fail to capture the long-term communication-sensing tradeoff.

\section{Conclusion}
This paper investigated an air-ground ISAC system supported by a UAV swarm with cross-region cooperation. A service-driven regional partitioning scheme was designed to organize UAV
service responsibility under imbalanced communication demand. In addition, an adaptive handshaking mechanism was introduced to characterize the trade-off between cooperative sensing performance and synchronization energy cost. Subsequently, a region-level multi-agent learning framework was developed to jointly optimize the UAV motion and transmission resources. Simulation results demonstrated the effectiveness of the proposed method, achieving a communication QoS of approximately 90\% and reducing the CRB by about 45\% compared to conventional baselines. In future work, we will further investigate dynamic regional reconfiguration and flexible UAV-swarm cooperation in air-ground ISAC scenarios.

\end{document}